\documentstyle[epsfig, rotate]{aipproc}

\begin{document}

\title{ Emission and Cooling Processes in a Hybrid Thermal-Nonthermal Plasma}
\author{ D. Lin and E. P. Liang }

\address{ Department of Space Physics and Astronomy, Rice University, Houston, TX 77005-1892 }

\maketitle

\begin{abstract} 
In a hybrid thermal-nonthermal plasma, we find that the dominant emission and absorption mechanisms are synchrotron by  nonthermal electrons and bremsstrahlung by thermal electrons. These two processes significantly change  the spectrum from inverse Compton scatterings at low energies. We also find that Coulomb collisions are effective in cooling down the lower energy electrons but do not significantly alter the emission pattern. Compton cooling is more effective in changing emission and absorption coefficients when the photon energy density is high. 

\end{abstract}
\section*{Introduction}
The accurate positioning of GRBs by BeppoSAX has allowed observation of gamma ray bursts over a wide range of wavelengthes. Spectra from these observations provide crucial tests of different emission models, one of which is the inverse Compton scattering (ICS) model [1]. The ICS model explains from first principles most of the known spectral evolution properties of GRBs [1,2]. Under the ICS model, GRB spectra suggest that the sources are thermal-nonthermal hybrid plasma [1,2] whose electron distribution is described as

$$N( \gamma ) =\left \{ \begin{array}{ll}
f_{th}( \gamma ) = A  \gamma  \sqrt{\gamma^{2}-1}~e^{-\frac{( \gamma-1 )}{T_{e}}}
& \gamma \leq \gamma_{th}\\ 
f_{nth}( \gamma )= B  \gamma^{-p}  
& \gamma \geq \gamma_{th}
\end{array} \right.  \eqno{(1)}$$
where A and B are determined by normalization and the continuity condition at 
$\gamma_{th}$.
This hybrid particle distribution has its own distinctive emission and absorption properties, which certainly affect the emerging spectra from the plasma. In this article, we will calculate the emission and absorption coefficients and incorporate them into the ICS model to make the model applicable to the multiwavelength spectra.

\begin{figure}[t] % fig
\centerline{
\psfig{figure=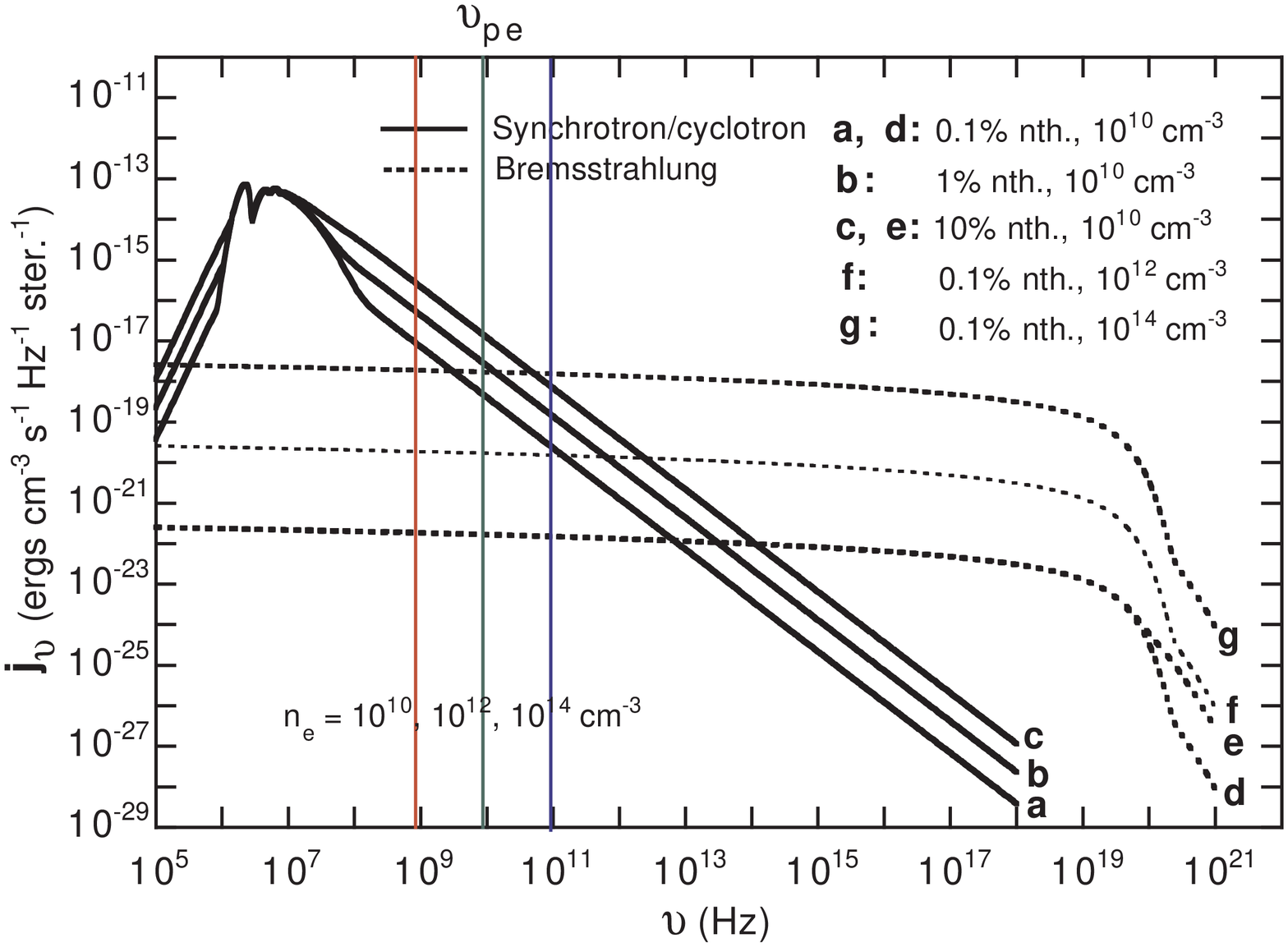, height=2.2in, width=5in}}
\caption{Emission coefficients of synchrotron/cyclotron and bremsstrahlung emissions,  where $\nu_{pe}$ is the plasma frequency. The emission coefficients  of cases f and g have been reduced by a factor of $10^2$ and $10^4$ respectively, so that they can be compared with cases a, b and c. For this calculation,  B = 1 Gauss and $T_e = 50$ keV.} 
\end{figure}
The significant roles of nonthermal electrons in 
emission and Compton scattering make it important to study the cooling processes of these electrons. Coulomb collisions and Compton scattering are the two processes we study. Aiming at modeling GRBs, we focus our calculation on the sources [1,2] with magnetic fields between 0.1 - 10 Gauss , $T_e $ less than 100keV, electron densities from $10^{10}$ - $10^{14} cm^{-3}$, and nonthermal fraction less than 10\%.

\section*{Emission And Absorption Coefficients }

Three major emission mechanisms in the hybrid source are cyclotron, synchrotron and bremsstrahlung emissions of electrons.
For an arbitrary normalized particle distribution function $f(\gamma)$, the emission coefficient of cyclotron/synchroton emission is [3] 
$$j_{\nu}^{cyc} = \frac{e^2wn_e}{2c} \sum_{n=1}^{\infty} \int_{\beta_{min}}^{\beta_{max}} \frac{f(\gamma)}{\gamma \sqrt{\gamma^2 - 1}} \big[ ( 1 - \frac{\gamma_{||}^2}{\gamma^2})\frac{J_{n}^{\prime}(\xi)}{\gamma_{||}} + \frac{( cos\phi - \beta_{||})^2} {sin^2\phi}J_{n}(\xi)\big] d\beta_{||} \eqno(2) $$
where $\gamma = \frac{\gamma_n}{1-\beta_{||} cos\phi} $, $\gamma_n = \frac{n
\nu_c}{\nu}$, $\nu_c$ is the cyclotron frequency, and 
$ \xi = \frac{\nu}{\nu_c}sin\phi\sqrt{ \gamma(1-\beta_{||}^2) - 1}$.
The integration limits in equation (1) are set to make $\xi$ a real number.

\begin{figure}[t] % fig
\centerline{
\psfig{figure=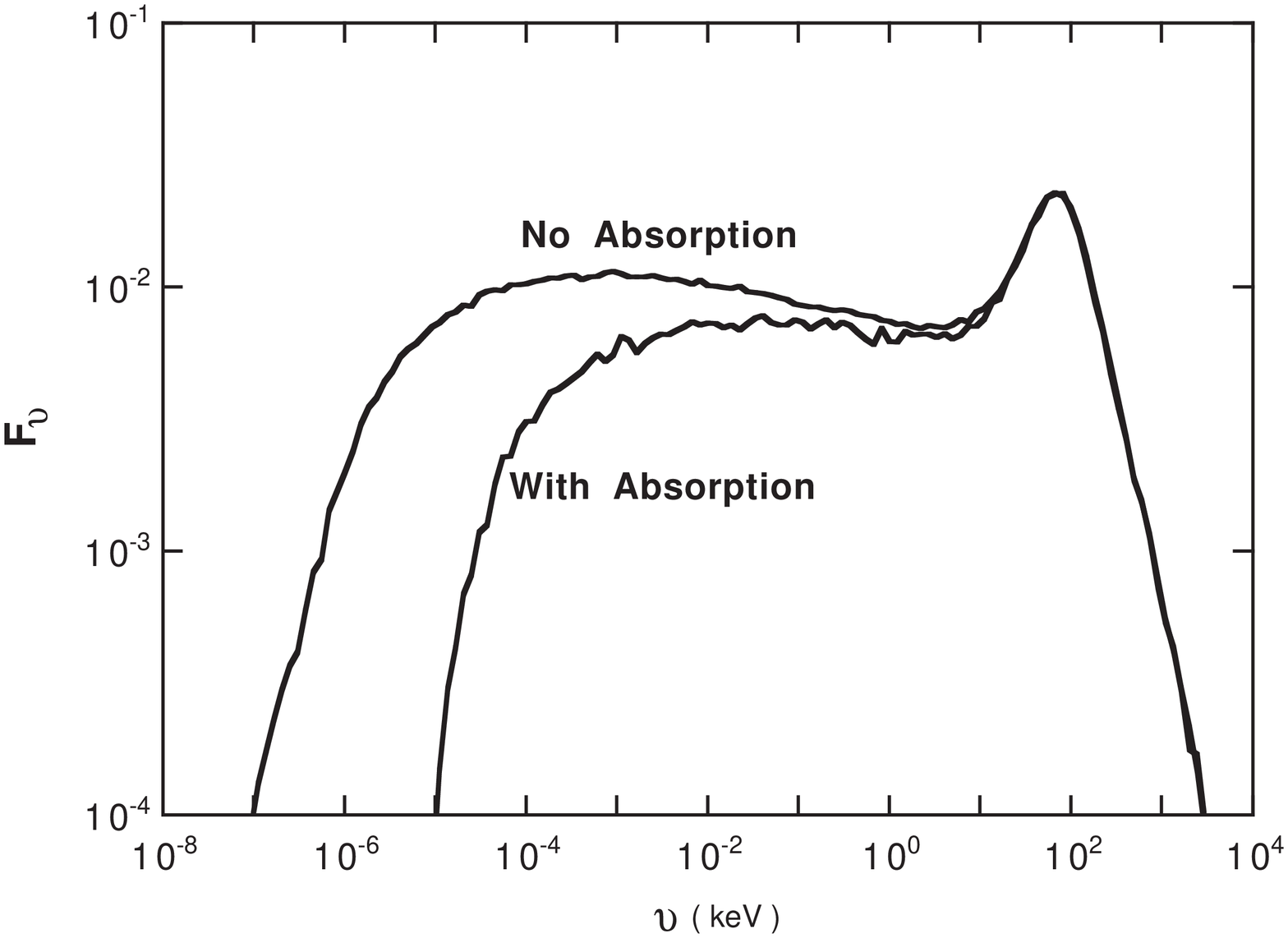, height=2in, width=5in}}
\caption{Spectra from ICS with and without absorption. The soft photon source with $T_r=5.11\times 10^{-7}keV$ sits in the center of a spherical plasma. For this calculation, B = 10 Gauss, $\tau_T$=10, $T_e$ = 20 keV, $n_e = 10^{10}cm^{-3}$, and nonthermal fraction = 10\% } 
\end{figure}

The emission coefficient of bremsstrahlung emission is [4]
$$ j_{\nu}^{ff} = \frac{4e^6n_e^2}{3\pi m_e^2c^4} \int_{1}^{\infty} \frac{\gamma f(\gamma)}{\sqrt{\gamma^2 - 1}} \big[ ln(\frac{2mc^2(\gamma^2-1)}{h\nu}) - \frac{\gamma^2 -1}{2\gamma^2}\big] d\gamma \eqno(3) $$
The absorption coefficients of both processes are calculated by
                         $\alpha_{\nu} = \frac{c^2}{8\pi h\nu^3} j_\nu^\prime $,
where $ j_\nu^\prime $ is obtained from eqn. (2) or (3) by replacing $f(\gamma)$ with $f^\prime(\gamma)$ [5]. $f^\prime(\gamma)$ is:
$$ f^\prime(\gamma) = \big( \frac{f(\gamma^*)}{\gamma^* \sqrt{{ \gamma^*}^2 - 1}} - \frac{f(\gamma)}{\gamma\sqrt{{ \gamma}^2 - 1}}\big) \gamma\sqrt{ \gamma^2 - 1}  \eqno(4) $$
where $\gamma^* = \gamma - \frac{h\nu}{mc^2}$.      
\vspace{0.1cm}
Calculation results ( see Fig.1 ) show that cyclotron emission can not propagate in the plasma, because its frequency is well below the plasma frequency. Nonthermal bremsstrahlung emission is also negligible. Nonthermal synchrotron and thermal bremsstrahlung are the dominant emission mechanisms. The same conclusion can be drawn for the absorption processes. Fig.2 shows how the absorption and emission processes change the spectra from ICS.  The spectrum is significantly cut-off  at lower energies.  This may be the reason why not many radio counterparts are detected [6].
\section*{ Cooling of Nonthermal electrons }
As shown above, emission spectra are largely dependent on the particle distribution. The cooling processes of nonthermal particles affects not only the soft photon emission but also the inverse Compton scattering. Coulomb and Compton coolings are the two dominant processes. Synchrotron cooling is neglected because the magnetic field energy density is much less than electron's kinetic energy density.
The cooling processes are governed by $\frac{d\gamma}{dt} = - \frac{\gamma}{\tau_c}$, where $\tau_c $ is the characteristic cooling time of both processes [5,7]. 
 
\vspace{0.1cm}
Calculation results ( see Fig.3 ) show that 
Coulomb thermalization quickly cools down low energy electrons. However, these less energetic electrons contribute little to synchrotron emission. The emission coefficient above the plasma frequency does not significantly change as thermalization occurs. Compton cooling, on the other hand, cools down high energy electrons first. Loss of the high energy electrons reduces significantly synchrotron emission when the photon energy density is high. 

\begin{figure}[t] % fig
\centerline{
\psfig{figure=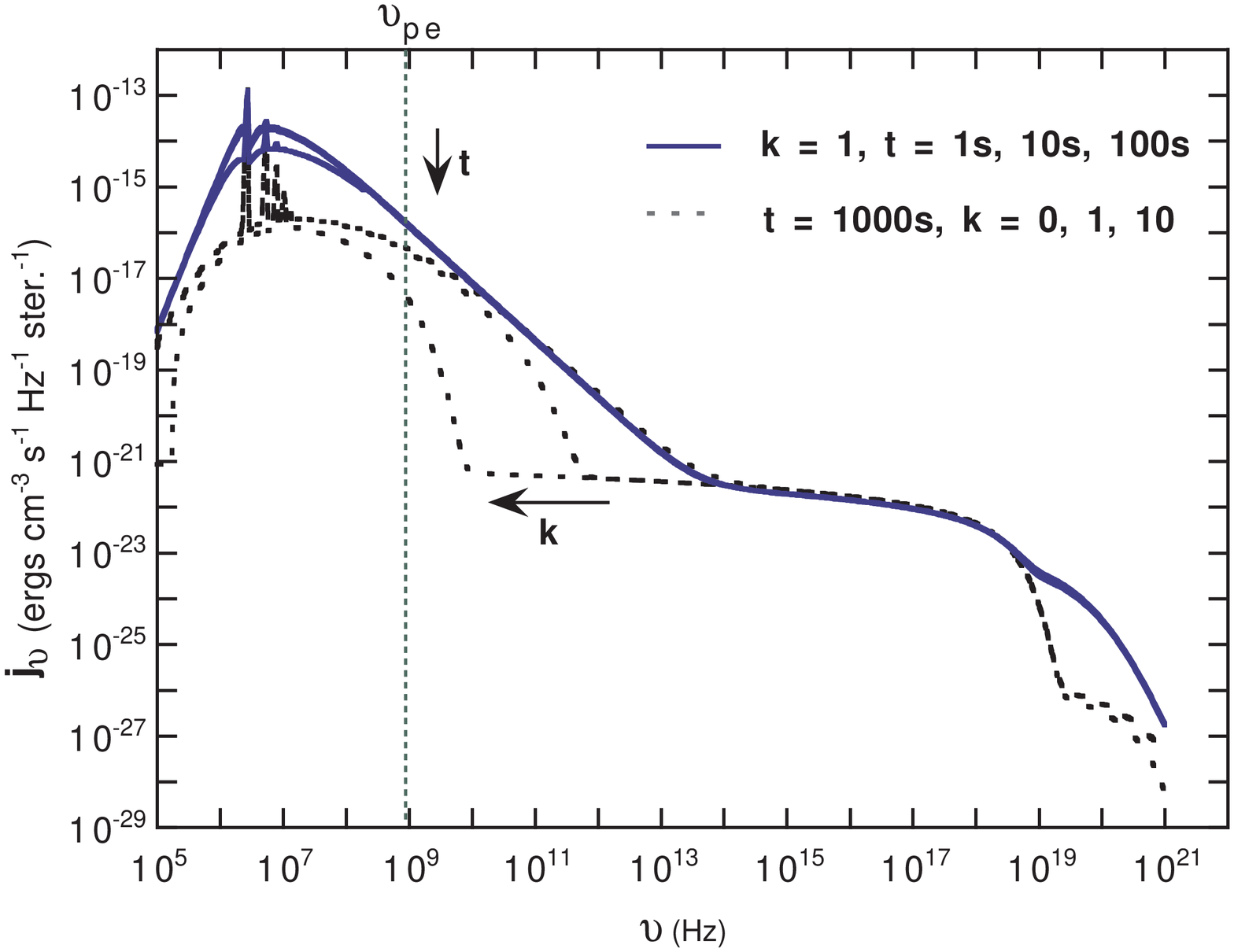, height=2.2in, width=5in}}
\caption{Effects of cooling processes on emission. Solid lines show the Coulomb collision effects, and dotted lines show Compton scattering effects. In this calculation, B = 1 Gauss, $n_e = 10^{10} cm^{-3}$, $T_e =$10 keV, $\gamma_{th} = 1.1$ and $k = \frac{U_{ph}}{n_ekT}$, where $U_{ph}$ is the photon energy density.} 
\end{figure}

\section*{Summary}
We have calculated the emission and absorption coefficients for a hybrid thermal-nonthermal plasma and find that nonthermal synchrotron and thermal bremsstrahlung are the dominant mechanisms. These two mechanisms have been incorporated into our inverse Compton scattering model so that it may be compared with radio observations of GRBs. Future multi-wavelength observations will be critical tests of this model. Our calculations also show that Coulomb thermalization does not significantly change the emission pattern but Compton cooling does.

\vspace{0.3cm}
{\bf Acknowledgements.} This work was partially supported by NASA grant NAG 5-3824.

\end{document}